\documentclass[preprint,aps,pra,showpacs,floatfix,superscriptaddress, 12pt]{revtex4-1}
\usepackage[utf8]{inputenc}
\usepackage[english]{babel}
\usepackage{geometry}
\usepackage{array}
\usepackage{siunitx}
\usepackage{graphicx}
\usepackage{amsmath}
\usepackage{amsfonts}
\usepackage{amssymb}
\usepackage{amsthm}
\usepackage{epsf}
\usepackage{bm}
\newcommand{\de}[1]{E^{(#1)}}

\usepackage{color}
\definecolor{BLUE}{rgb}{0.0,0.0,1.0}

\usepackage{soul}
\soulregister\cite7
\soulregister\ref7

\newcommand{\gf}{g_{\mathrm{free}}}
\newcommand{\gbb}{g^{(2)}}

\newcommand{\exstm}{${}^2 {P}_{3/2}~$}
\newcommand{\grstm}{${}^2 {P}_{1/2}~$}

\newcommand{\gbblo}{ g^{(2)}_{\mathrm{lo}}}

\newcommand{\dgbbqed}{\Delta g^{(2)}_{\mathrm{QED}}}

\newcommand{\dgbbintz}{\Delta g^{(2)}_{\mathrm{1ph}}}

\newcommand{\faz}{F(\alpha Z)}

\newcommand{\ha}{H_1^{\mathrm{rad}}}
\newcommand{\hb}{H_2^{\mathrm{rad}}}

\newcommand{\balpha}{{\mbox{\boldmath$\alpha$}}}
\newcommand{\bnabla}{{\mbox{\boldmath$\nabla$}}}

\newcommand{\be}{\begin{eqnarray}}
\newcommand{\ee}{\end{eqnarray}}

\newcommand{\bfr}{{\bf r}}

\newcommand{\matrixel}[3]{\langle #1 | #2 | #3 \rangle}
\newcommand{\ArB}{${}^{40}\mathrm{Ar}^{13+}$}
\newcommand{\SnB}{${}^{118}\mathrm{Sn}^{45+}$}

\newcommand{\gBB}[1]{g^{(2)}_{#1}}

\newcommand{\gBBint}{\Delta\gBB{\mathrm{int}}}

\newcommand{\mub}{\mu_\textrm{B}}

\newcommand{\Umag}{U_{\textrm{m}}}

\newcommand{\sumn}{\sideset{}{'}\sum\limits_{n}}
\newcommand{\summn}{\sideset{}{'}\sum\limits_{n_1,n_2}}

\newcommand{\dgq}[2]{\mbox{$\Delta g_{#1}^{(2 #2)}$}}

\newcommand{\nrlgbb}[1]{\mbox{$g_{{\mathrm{FS}}#1}^{(2)}$}}

\newcommand{\mel}[3]{\langle #1|#2|#3\rangle}

\AtBeginDocument{%

}
\AtBeginDocument{%

}
\begin{document}

\title{Quadratic Zeeman effect in light boron-like ions}

\author{V.~A.~Agababaev}
%\affiliation{Saint-Petersburg State Electrotechnical University ``LETI'', Saint-Petersburg, Russia}
\affiliation{ITMO University, St. Petersburg 197101 Russia}
\affiliation{HSE University, St. Petersburg 194100 Russia}
\author{A.~V.~Volotka}
\affiliation{ITMO University, St. Petersburg 197101 Russia}

\author{D.~A.~Glazov}
\affiliation{ITMO University, St. Petersburg 197101 Russia}

\affiliation{Petersburg Nuclear Physics Institute named by B. P. Konstantinov of
National Research Centre ``Kurchatov Institute'', Gatchina 188300 Russia}
\author{A.~V.~Malyshev}
\affiliation{Petersburg Nuclear Physics Institute named by B. P. Konstantinov of
National Research Centre ``Kurchatov Institute'', Gatchina 188300 Russia}
\affiliation{St. Petersburg State University, St. Petersburg 199034 Russia}
\author{M.~M.~Osiptsov}
\affiliation{ITMO University, St. Petersburg 197101 Russia}

\author{V.~M.~Shabaev}
\affiliation{Petersburg Nuclear Physics Institute named by B. P. Konstantinov of
National Research Centre ``Kurchatov Institute'', Gatchina 188300 Russia}
\affiliation{St. Petersburg State University, St. Petersburg 199034 Russia}

\begin{abstract}
The quadratic Zeeman effect is calculated for the ground \grstm state of light boron-like ions in the range of nuclear-charge numbers $Z = 10$--$24$.
The calculations are performed in the Furry picture using three models for the zeroth-order approximation potential: pure nuclear Coulomb potential and two effective screening potentials --- core-Hartree and Kohn-Sham.
First-order perturbation-theory contributions are considered: the one-photon-exchange correction and the radiative corrections associated with the self-energy and vacuum-polarization diagrams.
The dominant contributions from the self-energy diagrams are calculated within a rigorous QED approach.
The vacuum polarization corrections are obtained within the electric-loop approximation in the leading order, which is given by the Uehling potential.
As a result, theoretical predictions for the contribution of the quadratic Zeeman effect to the binding energy of the valence electron in the \grstm state are obtained. The results can be used for the analysis of high-precision $g$-factor and fine-structure splitting measurements in boron-like highly charged ions.
\end{abstract}
%
%%%%%%%%%%%%%%%%%%%%%%%%%%%%%%%%%%%%%%%%%%%%%%%%%%%%%%%%%%%%%%%%%%%%%%%%
% 
\maketitle
%\hspace{\Parindent}
\section{Introduction}

Studies of the Zeeman effect have significantly advanced over the past two decades. Modern high-precision measurements of the $g$-factor include hydrogen-, lithium-, and boron-like ions up to the heaviest elements, such as tin \cite{Morgner:23:nature, Morgner:2025:PRL, Morgner:2025:Science}, achieving accuracy at the level of less than one part per billion \cite{haeffner:00:prl, verdu:04:prl, sturm:11:prl, sturm:13:pra, wagner:13:prl, Arapoglou:2019:PRL, glazov:19:prl, Heisse:2023:PRL, Spiess:2025:PRL}, with the record values of statistical uncertainty being of the order of $4\times 10^{-11}$ \cite{sturm:13:pra}.

A landmark achievement in this field is the most precise determination of the electron mass to date, obtained through combined theoretical and experimental studies of the $g$-factor of light hydrogen-like ions \cite{sturm:14:n,CODATA:2025:025002}. Investigations of hydrogen- and lithium-like systems have also enabled a rigorous test of relativistic bound-state QED methods in the presence of a magnetic field \cite{wagner:13:prl,volotka:14:prl,koehler:16:nc,yerokhin:17:pra,shabaev:17:prl}. In the future, it is expected that $g$-factor measurements of highly charged ions will allow an independent determination of the fine-structure constant $\alpha$ \cite{shabaev:06:prl,volotka:14:prl-np,yerokhin:16:prl}, the determination of nuclear parameters \cite{Quint:2008:032517, yerokhin:11:prl, Shabaev:2022:prl, Sailer:2022:nature}, and searches for new physics \cite{Sailer:2022:nature, Debierre:2020:PLB, Debierre:2022:PRA, Akulov:2025:52}.

The quadratic Zeeman effect has been the subject of active scientific investigations for nearly a century, beginning with the first observations by Jenkins and Segr\'e \cite{jenkins:pr:1939} and the subsequent theoretical description by Schiff and Snyder \cite{shiff:pr:1939}. Interest in it is partly due to the observation of astrophysical objects such as magnetars, where magnetic fields reach values of  $10^{11}$ T, making the nonlinear Zeeman contribution dominant \cite{preston:1970, Kemic:1975, Moran:1998}. At the same time, this effect is studied in solids, Bose–Einstein condensates \cite{Stenger:98:n, Thilderkvist:94}, as well as in atoms, molecules, and exotic systems such as positronium \cite{garton:1969, feinberg:pra:1990, Raoult:2005, numazaki:pra:2010}.

However, research on the quadratic Zeeman effect in highly charged ions is still fragmentary and, in fact, is only in its initial phase.
For the $1s$ state, nonlinear effects are negligible in laboratory-strength fields. However, for excited states, they can be significantly enhanced due to mixing of fine-structure levels. This occurs, in particular, for $P$ states in lithiumlike \cite{zinenko:25} and boronlike \cite{lindenfels:13:pra} ions. High-precision measurements of the ground-state $g$-factor of boronlike ions have been performed by the ALPHATRAP collaboration for argon \ArB and tin \SnB \cite{Arapoglou:2019:PRL,Morgner:2025:PRL}. For the excited \exstm state of \ArB, the $g$-factor was first measured in Ref.~\cite{Egl:2019:PRL} and subsequently significantly improved using quantum logic methods \cite{Micke:2020:Nature}. In all these experiments, the quadratic-in-field contribution had a noticeable impact on the transition frequencies and required careful treatment for the interpretation of the results. An even more precise value of the $g$-factor of the \exstm state in \ArB was obtained in a recent experiment aimed at developing a precision atomic clock based on $^{40}\textrm{Ar}^{13+}$ \cite{King:2022:nature}. Although the magnetic field in that experiment was too weak to observe nonlinear effects, further improvement in the accuracy of the $g$-factor determination will require increasing the field strength to about one tesla, which will give rise to significant quadratic and cubic contributions. Additionally, the second-order contribution in the magnetic field has recently been measured for the $^3P_0 \to {}^3P_1$ transition in carbon-like calcium \cite{Spiess:2025:PRL}.

%The most recent measurements of the $g$ factor of the ground \grstm state in boron-like ions were carried out by the ALPHATRAP collaboration for argon \ArB and tin \SnB \cite{Arapoglou:2019:PRL, Morgner:2025:PRL}. For the excited \exstm state, the $g$ factor was first measured in the \ArB ion in Ref. \cite{Egl:2019:PRL} and was subsequently significantly refined using the quantum-logic spectroscopy \cite{Micke:2020:Nature}. In a recent experiment aimed at developing precision atomic clocks based on $^{40}\textrm{Ar}^{13+}$, the most accurate value to date for the $g$ factor of the excited state of boron-like argon was obtained \cite{King:2022:nature}. This work relies on the analysis of the fine-structure transition frequency in a magnetic field. The quadratic Zeeman effect shifts the energy levels, thereby potentially affecting the measurement results. Although in this experiment the magnetic field was too weak to observe this effect, increasing the magnetic field up to a value of the order of 1 T would make this effect significant. It should be noted that such an increase in the external field would also significantly enhance the precision of the $g$-factor determination.

In the theoretical studies of the $g$ factor of the ground state of boron-like argon \ArB, several high-precision methods were used to take into account the interelectronic interaction, including the CI-DFS \cite{glazov:13:ps, shchepetnov:15:jpcs, cakir:20:pra}, GRASP2K \cite{verdebout:14:adndt}, MCDFME \cite{marques:16:pra} methods, perturbation theory \cite{shchepetnov:15:jpcs, agababaev:18:jpcs}, and the coupled-cluster method \cite{Maison:19:pra}. 
In addition, within the framework of a rigorous QED approach, first-order self-energy and vacuum-polarization corrections  were considered \cite{agababaev:18:jpcs, agababaev:19:XRS, cakir:20:pra}, and calculations of the nuclear-recoil effect were performed \cite{glazov:18:os, aleksandrov:18:pra, Malyshev:2020:OS, Glazov:2020:PRA}.
A comprehensive review of modern theoretical approaches to the $g$ factor in lithium and boron-like highly charged ions is presented in, e.g., Ref. \cite{Glazov:2023:atoms}.

Theoretical studies of the quadratic Zeeman effect in highly charged ions have already been carried out by our group \cite{Agababaev:17:NIMB, agababaev:18:jpcs, zinenko:25}, including the one-photon exchange corrections \cite{varentsova:18:pra} and one-loop QED corrections \cite{Agababaev:2025:PRA}. Within the same approaches, the cubic Zeeman effect was also studied \cite{Varentsova:17:NIMB, varentsova:18:pra, zinenko:25}, as well as nuclear magnetic shielding \cite{Volchkova:17:NIMB, Volchkova:2020:arxiv}. Theoretical estimates of the energy-level shifts induced by the nonlinear Zeeman effects proved to be essential for interpreting the results of $g$-factor measurements in boron-like argon \ArB \cite{Arapoglou:2019:PRL, Egl:2019:PRL}.

In the present work, we present calculations of the quadratic Zeeman effect for the \grstm state of light boron-like ions in the nuclear charge range $Z = 10$--$24$. We compute the corresponding contribution to the binding energy of the valence $2p_{1/2}$ electron, taking into account the one-photon exchange and one-loop QED corrections.

We use relativistic units, in which $\hbar=c=m_e=1$, and Heaviside charge units, where the fine-structure constant $\alpha = e^2/(4\pi)$, with the electron charge $e < 0$.

\section{Theory}

We consider a boron-like ion in the ground $(1s)^2(2s)^2 2p\,^2P_{1/2}$ state. The nucleus is assumed to be spinless. The system is placed in a weak homogeneous magnetic field $\mathbf{B}$, oriented along the $z$-axis.

The energy of a many-electron state $|A\rangle$ of an atom placed in a magnetic field can be expanded in a  series 
\begin{equation}
\label{eq:pt1}
E = E^{(0)} + E^{(1)} + E^{(2)} + \cdots,
\end{equation}
where $E^{(0)}$ is the energy of the state $|A\rangle$ in the absence of the field, $E^{(1)}$ is the first-order contribution in magnetic field, $E^{(2)}$ is the second-order contribution, etc.
The first-order contribution is proportional to the $z$-axis projection $M_J$ of the total angular momentum $J$ of the state $|A\rangle$, and it can be conveniently expressed it in terms of the dimensionless $g$ factor:
\begin{equation}
\de{1} = \mub B M_J g \,,
\end{equation}
where $\mub = |e|/2$ is the Bohr magneton.
In turn, the second-order contribution in equation \eqref{eq:pt1} can be written as
\begin{equation}
\label{eq:ebb}
\de{2} = (\mu_B B)^2 \gbb(M_J)\,.
\end{equation}
The dependence of $\gbb$ on $M_J$ is not linear, but represents an even function of $M_J$, which has important physical consequences. For states with total angular momentum $J = 1/2$, the quadratic effect is the same for both $M_J$ sublevels and does not contribute to the Zeeman splitting. 
However, for states with $J > 1/2$, the quadratic shift differs for sublevels with different values of $|M_J|$ and breaks the equidistance of the Zeeman splitting. In the state $^2P_{1/2}$ studied in this work, the quantum numbers $J$ and $M_J$ are completely determined by the corresponding quantities $j_a$ and $M_a$ of the valence $2p_{1/2}$ electron.

We also note that we consider the contribution of the quadratic Zeeman effect to the binding energy of the valence electron, which can be obtained as the difference of the binding energies of the states $(1s)^2(2s)^2 2p\,^2P_{1/2}$ and $(1s)^2(2s)^2 \,^1S_{0}$. Within the framework of perturbation theory, this means that we omit all corrections that do not involve the state of the valence electron, but correspond only to the states of the closed shells.

The interaction with the magnetic field $\mathbf{B}$ is conventionally introduced by means of the minimal substitution for the canonical momentum: $\mathbf{p} \rightarrow \mathbf{p} - e\mathbf{A}_{\rm cl}$, where the classical vector potential $\mathbf{A}_{\rm cl}$ is related to the magnetic field by the expression $\mathbf{B} =[ \bnabla\times \mathbf{A}_{\rm cl}]$. For a homogeneous magnetic field $\mathbf{B} = B \mathbf{e}_z$, we use the symmetric gauge, in which $\mathbf{A}_{\rm cl} = \tfrac{1}{2}[\mathbf{B}\times\mathbf{r}]$.
The simplest estimate for the quadratic Zeeman coefficient $\gbb$ can be obtained from the Breit-Pauli equation for a valence electron in an external field $\mathbf{A}_{\rm cl}$, which corresponds to the one-electron approximation:
\begin{equation}
\label{eq:nrl0}
    {H}_{\rm BP} = \frac{(\mathbf{p} - e\mathbf{A}_{\mathrm{cl}})^2}{2} + V_{\text{nuc}}(r) - \boldsymbol{\mu}\cdot\mathbf{B} + {H}_{\rm rel} \,.
\end{equation}
Here $V_{\text{nuc}}(r)$ is the nuclear potential, $r=|\mathbf{r}|$, ${\bm\mu}= -2\mu_B{\bf s}$ is the magnetic moment of the electron, proportional to the spin ${\bf s}$ (where we neglect the anomalous magnetic moment of the electron), and ${H}_{\rm rel}$ is the operator that includes all relativistic corrections, among which for the present consideration the most significant is the spin-orbit interaction operator ${H}_{\rm SO} \sim\mathbf{l}\cdot\mathbf{s}$, which determines in the leading order $(\alpha Z)^4$ the fine-structure splitting $2p_{1/2}$--$2p_{3/2}$.
Using the symmetric gauge, we obtain
\begin{equation}
 H_{\rm BP} = \frac{{\bf p}^2}{2} + V_{\rm nuc}(r) + \mu_B B\,V_1 + (\mu_B B)^2 V_2 + {H}_{\rm rel}\,.
\end{equation}
The operators for the first and second orders in the magnetic field are given by the expressions:
\begin{align}
\label{eq:nrl1}
V_1 &= l_z + 2s_z = j_z + s_z\,, \\
\label{eq:nrl2}
V_2 &= \frac12\bigl(r^2 - z^2\bigr)\,. 
\end{align}
Here $l_z$, $s_z = \sigma_z/2$, and $j_z = l_z + s_z$ are the projections of the orbital, spin, and total angular momenta.

The discussed estimate of the quantity $\gbb$ includes two contributions:
\begin{equation}
\label{eq:g_lo_BP}
\gbblo = g^{(2)}_{\mathrm{lo}-1} + g^{(2)}_{\mathrm{lo}-2}
\end{equation}
which arise, respectively, from the second-order perturbation-theory correction for the operator $V_1$ and from the first-order perturbation-theory correction for the operator $V_2$:
\begin{align}
g^{(2)}_{\mathrm{lo}-1} &=\sumn
\frac{\langle \bar a|V_1|\bar n\rangle\langle \bar n|V_1|\bar a\rangle}
     {\bar \varepsilon_a-\bar \varepsilon_n}, \label{eq:nrl6} \\
g^{(2)}_{\mathrm{lo}-2} &= \langle \bar a|V_2|\bar a\rangle. \label{eq:nrl6:1}
\end{align}
In these expressions, $|\bar a\rangle = |\overline{2p_{1/2}}\rangle$, $|\bar n\rangle$ denote the two-component eigenfunctions of the operator \eqref{eq:nrl0} in the absence of a magnetic field, and $\bar \varepsilon_a$, $\bar \varepsilon_n$ are the corresponding eigenvalues.
The summation in (\ref{eq:nrl6:1}) runs over the complete spectrum; the prime on the summation sign here and in what follows means that terms with vanishing denominators should be omitted.

Let us consider the contribution of the leading order in $\alpha Z$ to (\ref{eq:g_lo_BP}).
Note that the sum in \eqref{eq:nrl6} is mainly determined by the term with the intermediate state $|\bar b\rangle \equiv |\overline{2p_{3/2}}\rangle$, which together with the studied state $2p_{1/2}$ forms the fine structure in the spectrum of a hydrogen-like ion. Indeed, the corresponding denominator $\bar \varepsilon_a - \bar \varepsilon_b \equiv -\Delta E_{\rm FS}$ is of the order $(\alpha Z)^4$, while for all other states the estimate $(\alpha Z)^2$ holds. Moreover, the matrix element $\langle \bar a | V_1 | \bar n \rangle$ in the numerator is of the order $(\alpha Z)^0$ for $|\bar n\rangle = |\bar b\rangle$ and $(\alpha Z)^2$ for $|\bar n\rangle \neq |\bar b\rangle$. It can also be shown that the contribution $g^{(2)}_{\mathrm{lo}-2}$ is of the order $(\alpha Z)^2$, so it can be omitted. Keeping the dominant term in equation \eqref{eq:nrl6}, we obtain
\begin{align}
\nrlgbb{}{} 
 &=  \frac{\langle \bar a|V_1|\bar b\rangle\langle \bar b|V_1|\bar a\rangle}
          {\bar \varepsilon_a-\bar \varepsilon_b} 
% = -\frac{\bigl|\langle \Omega_{\kappa_a M_a}|\sigma_z/2|\Omega_{\kappa_b M_b}\rangle\bigr|^2}
%           {\Delta E_{\rm FS}}
 \approx -\frac{(3/2)^2 - M_a^2}{9\,\Delta E_{\rm FS}}\,, \label{eq:17-1}
\end{align}
where we have changed the notation from $g^{(2)}_{\mathrm{lo}}$ to $g^{(2)}_{\mathrm{FS}}$ to emphasize the approximation we made, and also used the fact that in the leading order in $\alpha Z$ the radial functions of the states $|\overline{2p_{1/2}}\rangle$ and $|\overline{2p_{3/2}}\rangle$ coincide.
Substituting $M_a \equiv M_J = \pm 1/2$, we obtain
\begin{equation}
\label{eq:nrl8}
\nrlgbb{}{} \approx -\frac{2}{9\,\Delta E_{\rm FS}}\,.
\end{equation}
In the case of $V_{\text{nuc}}(r) = -\alpha Z/r$, one can use the Sommerfeld formula to estimate $\Delta E_{\rm FS}$ in the leading order $(\alpha Z)^4$, which finally leads to the expression
\begin{equation}
\label{eq:nrl:gbb}
\nrlgbb{} \approx - \frac{64}{9(\alpha Z)^4}\,.
\end{equation}
It is well known that in highly charged ions the fine structure interval $\Delta E_{\text{FS}}$ strongly depends on correlation effects. When treating them within perturbation theory, a widely used approach consists of adding a local screening potential $V_{\text{scr}}(r)$ to the unperturbed Hamiltonian. The same potential is then subtracted from the interaction operator, which leads to a rearrangement of perturbation-theory series, accelerating their convergence. The described method allows one to partially take into account the effects of interelectronic interaction already in the zeroth approximation.
In this regard, formula (\ref{eq:nrl8}) has a broader applicability, since one can substitute into it estimates for $\Delta E_{\text{FS}}$ obtained in the leading order for various potentials $V_{\text{scr}}(r)$.

A rigorous QED theory of the quadratic Zeeman effect can be formulated within the framework of the two-time Green's function method \cite{Shabaev:02:pr}. In this approach, the effects of interelectronic interaction as well as the QED effects are taken into account by perturbation theory, so in the zeroth-order approximation the one-electron states are completely determined by the Dirac Hamiltonian:
\begin{equation}
\label{eq:dirac}
 H_{\rm D} = \boldsymbol{\alpha} \cdot \mathbf{p} + \beta  + V_{\mathrm{nuc}}(r)+V_{\mathrm{scr}}(r) \,.
\end{equation}
Here $\boldsymbol{\alpha}$, $\beta$ are the Dirac matrices.
As the screening potential $V_{\text{scr}}(r)$, in this work we use the core-Hartree (CH) and Kohn-Sham (KS) potentials \cite{pot:KS}.
These potentials can be expressed in terms of the charge density of the closed-shell electrons,
\begin{align}
\label{eq:dense_c}
 \rho_{\rm core}(r) = 2 \Big[ G_{1s}^2(r) + F_{1s}^2(r) \Big] 
+ 2 \Big[ G_{2s}^2(r) + F_{2s}^2(r) \Big] \, ,
\end{align}
and of all electrons of the ion,
\begin{align}
\label{eq:dense_t}
\rho_{\rm tot}(r) = \rho_{\rm core}(r) +\Big[ G_{2p_{1/2}}^2(r) + F_{2p_{1/2}}^2(r)\Big]  \, .
\end{align}
In (\ref{eq:dense_c}) and (\ref{eq:dense_t}), the functions $G_n$ and $F_n$ are the large and small components of the solutions of the Dirac equation, which are normalized according to $\int (G_n^2+F_n^2)\,dr=1$. They are determined self-consistently for the potentials
\begin{align}
\label{eq:V_CH}
 V_{\text{scr}}^{\rm CH}(r) = \alpha\int\limits_0^\infty \! dr' \, \frac{\rho_{\rm core}(r')}{{\rm max}\{r,r'\}} \,,
\end{align}
and
\begin{align}
\label{eq:V_KS}
V_{\text{scr}}^{\rm KS}(r) = 
\alpha \int\limits_0^\infty \! dr' \, \frac{\rho_{\rm tot}(r')}{{\rm max}\{r,r'\}}  -\frac{2\alpha}{3r} \left[ \frac{81}{32\pi^2} r \rho_{\rm tot}(r) \right]^{1/3} \,, 
\end{align}
respectively.

The replacement $\mathbf{p} \rightarrow \mathbf{p} - e\mathbf{A}_{\rm cl}$ in (\ref{eq:dirac}) allows one to obtain an expression for the operator descibing the interation with an external homogeneous magnetic field
\begin{equation}
\label{eq:Vm}
V_{\rm m} =  \mub B \,[\mathbf{r}\times \boldsymbol{\alpha}]_z\equiv ( \mub B) \Umag \,.
\end{equation}
For example, for the leading-order contribution, the method \cite{Shabaev:02:pr} yields:
\begin{equation}
\label{eq:gbb-dirac}
\gbblo = \sumn \frac{\mel{a}{\Umag}{n}\mel{n}{\Umag}{a}}{\varepsilon_a - \varepsilon_{n}} \,,
\end{equation}
where $|a\rangle = |2p_{1/2}\rangle$, $|n\rangle$ denote the four-component eigenfunctions of the operator \eqref{eq:dirac}, $\varepsilon_a$, $\varepsilon_n$ are the corresponding eigenvalues, and the summation extends also to the negative-energy spectrum.
Reasoning analogously to how it was done in (\ref{eq:nrl6}), it can be shown that the dominant contribution to (\ref{eq:gbb-dirac}) comes from the intermediate state $|b\rangle \equiv |2p_{3/2}\rangle$, separated from $|a\rangle$ by the fine structure:
\begin{equation}
    \nrlgbb{}{} 
 =  \frac{\langle a|\Umag|b\rangle\langle b|\Umag|a\rangle}
          {\varepsilon_a-\varepsilon_b} \label{eq:est-gbb} \,.
\end{equation}
Expression (\ref{eq:est-gbb}) in the leading order in $\alpha Z$ also leads to the formulas (\ref{eq:nrl8}) and (\ref{eq:nrl:gbb}).

Within the framework of QED perturbation theory, the expression for $\gbb$ can be represented in the form:
\begin{equation}
\label{eq:gbb}
 \gbb =  \gbblo + \gBBint + \dgbbqed\,,
\end{equation}
where $\gbblo$ is the leading-order contribution (\ref{eq:gbb-dirac}), $\gBBint$ is the interelectronic-interaction correction, and $\dgbbqed$ is the radiative correction. In the calculations performed in this work, only the contribution $\gBBint$ goes beyond the one-electron approximation. We note that for the studied state $^2P_{1/2}$, the unperturbed wave function is described by a single Slater determinant constructed from the eigenfunctions of the operator (\ref{eq:dirac}).

\begin{figure}[ht]
\centering
\includegraphics[width=0.90\linewidth]{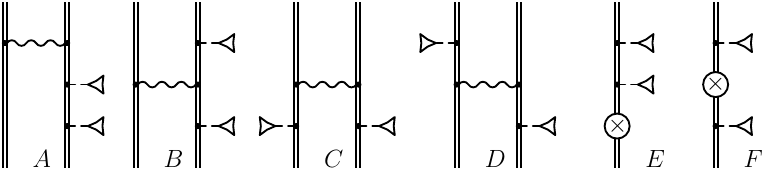}
\caption{One-photon-exchange corrections to the quadratic Zeeman effect. The double line denotes the electron propagator in the field $V_{\rm nuc}$ (or $V_{\rm nuc} + V_{\rm scr}$), the wavy line denotes the photon propagator, the dashed line ending with a triangle denotes the interaction with the magnetic field (\ref{eq:Vm}), the cross in a circle denotes the counterterm $\delta V = -V_{\rm scr}$}
\label{fig:1ph}
\end{figure}

As already noted above, the interelectronic interaction significantly affects the quadratic Zeeman effect, and the mechanism of this influence is twofold: first, it strongly changes the fine-structure splitting, which enters the denominators in expressions of the type of (\ref{eq:nrl8}) and (\ref{eq:est-gbb}); second, it affects the values of the matrix elements. Therefore, an accurate account of the correlation effects is extremely important.
For this reason, we perform the calculations not only for the nuclear Coulomb potential but also including the CH and KS screening potentials.
For a more careful treatment of the correlation contributions, we also consider the one-photon-exchange correction, which corresponds to the diagrams in Fig. \ref{fig:1ph}. The counterterm diagrams $E$ and $F$ in Fig. \ref{fig:1ph} are associated with the modification of the zeroth approximation, and they need to be taken into account only in calculations involving screening potentials. The formal expressions for the one-photon-exchange correction, obtained within the framework of a rigorous QED approach, are identical (up to symmetry coefficients) to those presented in Ref. \cite{Moskovkin:2008:OS}, provided the hyperfine interaction potential is replaced with the magnetic-field interaction potential.

\begin{figure}[ht]
\begin{center}
\hspace{-1cm}
\includegraphics[width=0.85\linewidth]{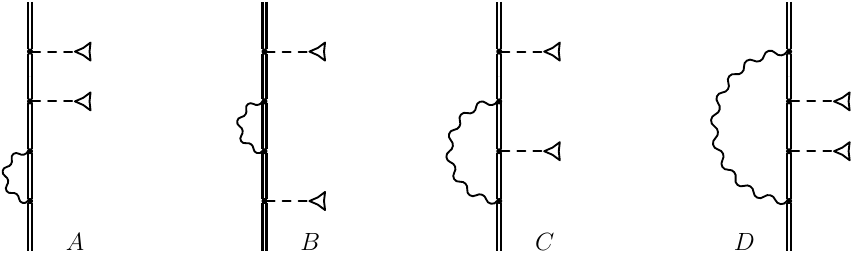}\
\caption{Self-energy corrections to the quadratic Zeeman effect. Notation as in Fig. \ref{fig:1ph}}
\label{fig:se}
\end{center}
\end{figure}

\begin{figure}[ht]
\begin{center}
\includegraphics[width=0.8\linewidth]{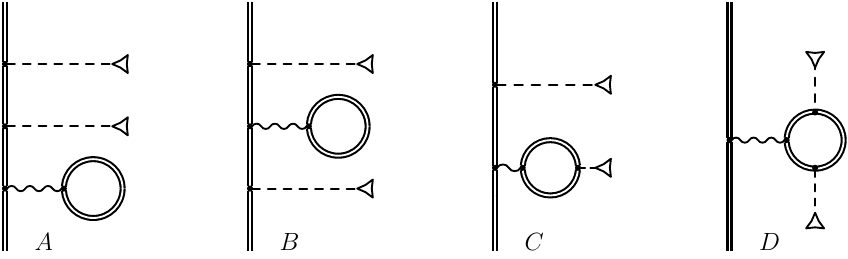}\
\caption{Vacuum-polarization corrections to the quadratic Zeeman effect. Notation as in Fig. \ref{fig:1ph}}
\label{fig:vp}
\end{center}
\end{figure}

The QED correction $\dgbbqed$ includes the self-energy (SE) and vacuum-polarization (VP) contributions:
\begin{equation}
 \dgq{\mathrm{QED}}{} = \dgq{\mathrm{SE}}{} + \dgq{\mathrm{VP}}{} \,,
\end{equation}
the diagrams for which are shown in Fig. \ref{fig:se} and Fig. \ref{fig:vp}, respectively.
These contributions have recently been examined in detail in our work \cite{Agababaev:2025:PRA}, where all the necessary formulas and calculation details can be found. We only note that the contribution of the self-energy diagrams is taken into account within the framework of the rigorous QED approach, while the contribution of the vacuum-polarization diagrams is considered in the electric-loop approximation, for which, in turn, the Uehling-potential is used. The rationale supporting the possibility of using these approximations is also given in \cite{Agababaev:2025:PRA}.

In Ref. \cite{Agababaev:2025:PRA} it was shown that the QED correction to the quadratic Zeeman effect for the $2p_{1/2}$ state can be approximately described using a pair of effective operators~\cite{Grotch:71:pra, Hegstrom:73:pra, glazov:04:pra}:
\begin{align}
\label{eq:ha}
 \ha &= \frac{\gf-2}{2} (\mu_{\rm B} B) \, \beta{\Sigma}_z \equiv  (\mu_{\rm B} B) U_1^{\rm rad}\,, \\
\label{eq:hb}
 \hb &= \frac{\gf-2}{2}\frac{\alpha Z}{2}(-i) \,\beta\frac{{\balpha}\cdot{\bfr}}{r^3} \, ,
\end{align}
where $\gf-2 \approx \alpha/ \pi$. We note that despite the wide application of these operators for the approximate treatment of QED contributions in the $g$-factor calculations, the correctness of their use for estimating similar corrections in the second order in the magnetic field is not obvious in advance. For example, in the case of $s$-states, their application leads to incorrect results \cite{Agababaev:2025:PRA}.

Constructing perturbation theory in the operators $V_{\rm m}$, $\ha$, and $\hb$ by the method \cite{Shabaev:02:pr}, we obtain that the QED correction to the coefficient $g^{(2)}$ for the $2p_{1/2}$ state can be approximately written as a sum of two terms,
\begin{align}
\label{eq:g_rad}
 \Delta g^{(2)}_{\rm rad} = \Delta g^{(2)}_{\rm rad-1} + \Delta g^{(2)}_{\rm rad-2} \, ,
\end{align}
where
\begin{align} \label{eq:g_rad_1}
 \Delta g^{(2)}_{\rm rad-1} &= 2 \sumn \frac{\matrixel{a}{\Umag}{n}\matrixel{n}{U_1^{\rm rad}}{a}}{\varepsilon_a - \varepsilon_n} \, , \\
\label{eq:g_rad_2}
 \Delta g^{(2)}_{\rm rad-2} &= \summn \bigg[ 2\,\frac{\matrixel{a}{\Umag}{n_1}\matrixel{n_1}{\Umag}{n_2}\matrixel{n_2}{\hb}{a}}{(\varepsilon_a - \varepsilon_{n_1})(\varepsilon_a - \varepsilon_{n_2})}+
 \frac{\matrixel{a}{\Umag}{n_1}\matrixel{n_1}{\hb}{n_2}\matrixel{n_2}{\Umag}{a}}{(\varepsilon_a - \varepsilon_{n_1})(\varepsilon_a - \varepsilon_{n_2})}\bigg] \nonumber \\
 &-\sumn\bigg[2\,\frac{\matrixel{a}{\Umag}{n}\matrixel{n}{\hb}{a}}{(\varepsilon_a - \varepsilon_n)^2}\matrixel{a}{\Umag}{a} + \frac{\matrixel{a}{\Umag}{n}\matrixel{n}{\Umag}{a}}{(\varepsilon_a - \varepsilon_n)^2}\matrixel{a}{\hb}{a}\bigg] \,.
\end{align}
Let us consider the leading contribution in $\alpha Z$ in expression (\ref{eq:g_rad}). As in the case of formulas (\ref{eq:g_lo_BP}) and (\ref{eq:gbb-dirac}), the dominant contribution comes from the intermediate states $2p_{3/2}$:
\begin{align}
 \Delta g^{(2)}_{\rm FS,rad-1} &= 2 \frac{\matrixel{a}{\Umag}{b}\matrixel{b}{U_1^{\rm rad}}{a}}{\varepsilon_a - \varepsilon_b} \, , \\
 \Delta g^{(2)}_{\rm FS,rad-2} &= \frac{g^{(2)}_{\rm FS}}{\varepsilon_a - \varepsilon_{b}} \left[ \matrixel{b}{\hb}{b} - \matrixel{a}{\hb}{a} \right] \,,
\end{align}
where we have again introduced the additional index ``FS'' and taken into account that the operator $\hb$ conserves the angular quantum numbers.
Using the virial relations \cite{Shabaev:1991:4479}, it can be shown that for $V_{\rm nuc}(r) = -\alpha Z/r$ the following relation holds
\begin{align}
 \langle b|U_1^{\rm rad}|a\rangle \approx (\gf-2)\langle b|\Umag|a\rangle \, ,
\end{align}
and therefore
\begin{align}
 \Delta g^{(2)}_{\rm FS,rad-1}  \approx 2 (\gf-2) g^{(2)}_{\rm FS} \,.
\end{align}
Furthermore, from the virial relations one can also obtain that
\begin{equation}
 \langle b|\hb|b\rangle - \langle a|\hb|a\rangle \approx  (\gf-2) \frac{(\alpha Z)^4}{32} \, ,
\end{equation}
and, therefore,
\begin{align}
 \Delta g^{(2)}_{\rm FS,rad-2}  \approx - (\gf-2) g^{(2)}_{\rm FS} \,.
\end{align}
Finally, we obtain that
\begin{align}
\label{eq:nr_qed_g2}
 \Delta g^{(2)}_{\rm FS,rad}  \approx (\gf-2) g^{(2)}_{\rm FS} \,.
\end{align}
%
%==================================================================

\section{Results and discussion}

In this section, we present the results of our calculations of the quadratic Zeeman effect for the ground $(1s)^2(2s)^2 2p,^2P_{1/2}$ state of boron-like ions, namely, the contribution of this effect to the binding energy of the valence $2p_{1/2}$ electron is considered. The results are presented in terms of the coefficient $\gbb$ defined in (\ref{eq:ebb}). The calculations have been performed for ions with the nuclear charges in the range $Z = 10$--$24$. To describe the nuclear-charge distribution, the standard two-parameter Fermi model was used.

The results for the quadratic Zeeman effect and the interelectronic-interaction corrections are collected in Table~\ref{tab:int}. Here and below, the results obtained in calculations with the Coulomb (Coul) potential of the nucleus and with the inclusion of the core-Hartree (CH) and Kohn-Sham (KS) screening potentials are shown separately. For the leading-order (LO) contribution, calculated according to (\ref{eq:gbb-dirac}), estimates \eqref{eq:nrl:gbb} and \eqref{eq:est-gbb} related to the fine structure (FS) are also given. Both estimates reproduce the full Dirac value (\ref{eq:gbb-dirac}) with good accuracy for small nuclear charges, while their accuracy gradually decreases with increasing $Z$. The one-photon-exchange corrections (1ph), calculated using the screening potentials, are systematically smaller than the corresponding values calculated in the pure Coulomb potential. This is because the screening potentials partially take into account the averaged field of the electrons even in the zeroth-order Hamiltonian. As a result, the perturbation theory series are rearranged, with both the leading-order terms and the corresponding one-photon-exchange corrections implicitly containing the higher-order contributions. The convergence of the results can be judged by examining the total values of the calculated contributions, which are given for each of the ions in the last row. From Table~\ref{tab:int}, it can be seen that the calculations using different screening potentials are in good agreement with each other: individual contributions may differ significantly, but the total values are close. The values obtained for the pure Coulomb potential differ significantly, especially for small $Z$: taking into account only the first-order (in $1/Z$) corrections  is clearly insufficient for a correct description of the interelectronic-interaction effects in this case.

The QED corrections are presented in Table \ref{tab:QED}. The self-energy (SE) contribution was calculated using the method outlined in our recent work \cite{Agababaev:2025:PRA}. The uncertainty indicated in the table is purely numerical in nature, and it is related to the analysis of the convergence of the results with respect to the size of the employed bases. The vacuum-polarization (VP) contribution was calculated in the electric-loop and Uehling-potential approximations. We do not provide an uncertainty for this correction, as this can only be done reliably after a complete evaluation is performed. However, we expect that the approximation used in this work gives the dominant contribution. We also note that the VP contribution is significantly smaller than the SE contribution. From Table \ref{tab:QED}, it can be seen that the results obtained for various screening potentials are in good agreement with each other. The difference from the pure Coulomb result is again related to the partial account of the interelectronic-interaction effects.

\begin{figure}[ht]
\begin{center}
\includegraphics[width=0.8\linewidth]{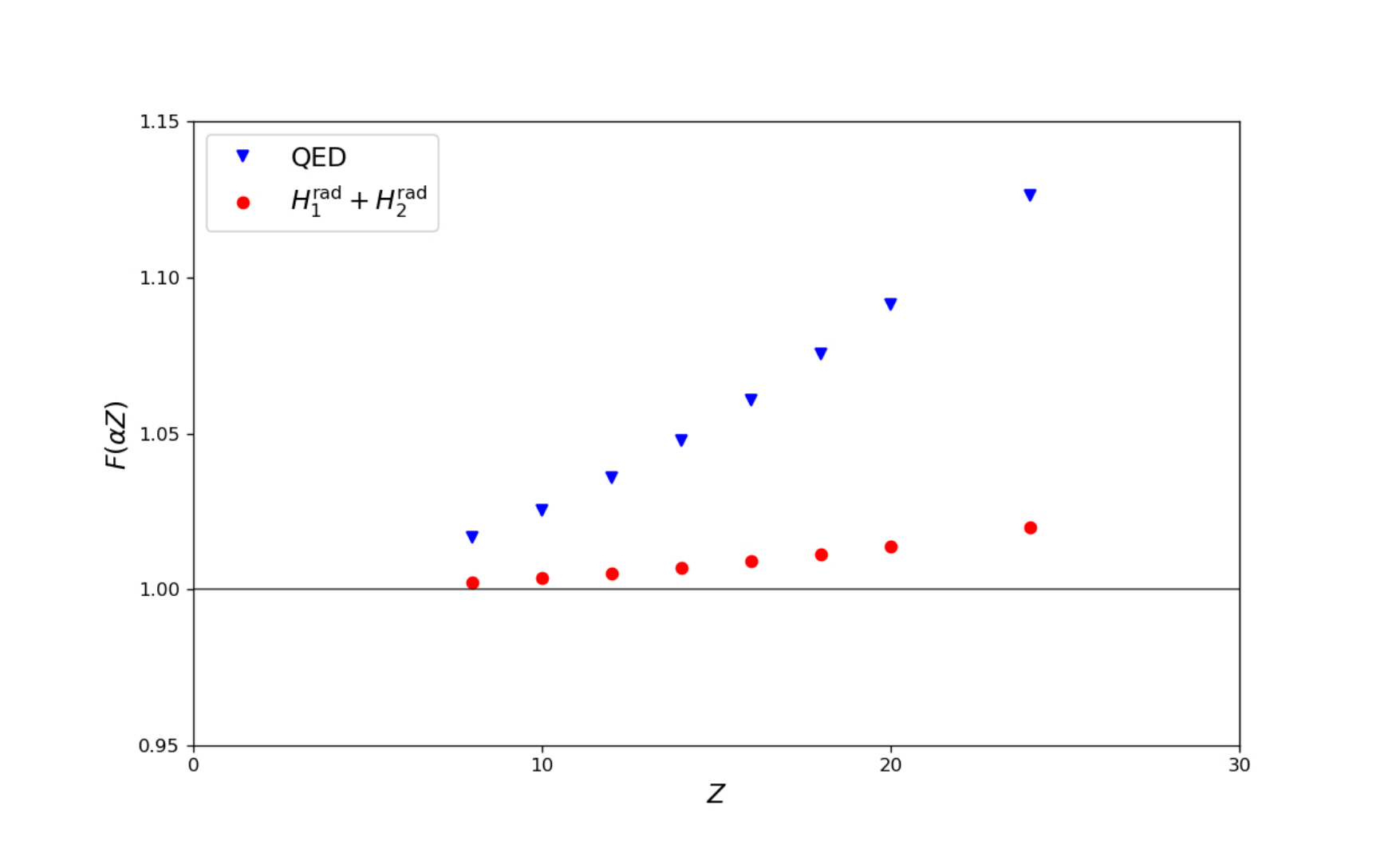}\
\caption{QED correction to the quadratic Zeeman effect for the $2p_{1/2}$ state in the region of small $Z$ in terms of the function $\faz$ defined in equation \eqref{eq:faz}. The rigorous QED correction obtained within the framework of the method \cite{Shabaev:02:pr} is shown by triangles, and the approximation obtained using the operators $\ha$ and $\hb$ is shown by circles. The leading-order contribution corresponds to the value $\faz \equiv 1$.}
\label{fig:qed}
\end{center}
\end{figure}

Motivated by the expression (\ref{eq:nr_qed_g2}), we also present the QED correction to the quadratic Zeeman effect for the $2p_{1/2}$ state via a function $F(\alpha Z)$, which is defined according to
\begin{equation}
\label{eq:faz}
 \dgbbqed = \frac{\alpha}{\pi}  \nrlgbb{} \faz \,,
\end{equation}
where $\nrlgbb{}$, in turn, is defined in (\ref{eq:est-gbb}).
We note that the function $\faz$ is defined here differently than in Ref. \cite{Agababaev:2025:PRA}.
According to (\ref{eq:nr_qed_g2}), the leading order corresponds to $\faz \equiv 1$.
In Table \ref{tab:QED}, the total QED correction in terms of $\faz$ is shown for each ion in the last row.
As can be seen, the function $\faz$ increases monotonically with increasing $Z$. At small $Z$, the correction is indeed close to unity. A comparison of the rigorous QED correction, obtained within the framework of the method \cite{Shabaev:02:pr}, with its approximation calculated using the operators $\ha$ and $\hb$ according to the formulas (\ref{eq:g_rad})--(\ref{eq:g_rad_2}), is presented in Fig. \ref{fig:qed}. To construct the graph, the results of the calculations with the Coulomb potential were used. It should be noted that in the region of small $Z$, the discrepancy between the exact QED correction and the approximation obtained using the effective operators does not exceed 3\%. However, with increasing $Z$, the difference increases, reaching about 5\% at $Z=24$.

The final results for the coefficient $\gbb$ are given in Table \ref{tab:gbb:final}. The final values represent the sum of three terms: the leading-order (LO) contribution $\gbblo$ from equation \eqref{eq:gbb-dirac}, the one-photon-exchange correction (1ph) $\dgbbintz$, and the QED correction $\dgbbqed$.
The difference between the results of calculations with the core-Hartree and Kohn-Sham screening potentials provides an estimate of the accuracy of the obtained theoretical predictions.
The significant magnitude of this uncertainty (especially for small $Z$) indicates that improving the accuracy will require a more rigorous account of interelectronic-interaction contributions. This implies both a systematic investigation using various screening potentials and the calculation of higher-order corrections in $1/Z$, which will be the subject of our further research.

\section{Summary}

In this work, the quadratic Zeeman effect for the ground  $(1s)^2(2s)^2 2p\,^2P_{1/2}$ state of boron-like ions has been considered in the range of nuclear charges $Z = 10$--$24$, and the energy shift of the valence $2p_{1/2}$ electron caused by it has been calculated. The employed approach systematically includes the first-order (in $1/Z$) interelectronic-interaction corrections and the one-loop QED corrections.
The calculations have been performed in the Furry picture both for the Coulomb potential of the nucleus and with the inclusion of the core-Hartree and Kohn-Sham screening potentials in the zeroth-order Hamiltonian.
The most accurate theoretical predictions for the contribution of the quadratic Zeeman effect have been obtained, which can be used for the analysis of high-precision experiments.
%
%%%%%%%%%%%%%%%%%%%%%%%%%%%%%%%%%%%%%%%%%%%%%%%%%%%%%%%%%%%%%%%%%%%%%%%%
%
%\ack
\section*{Acknowledgements}

The work was supported by the Ministry of Science and Higher Education of the Russian Federation (project No. FSER-2025–0012). %and the Russian Science Foundation (project No. 22-12-00258).

The work of A.~V.~M. was supported by the Foundation for the Development of Theoretical Physics and Mathematics "BASIS" (project No. 24-1-2-74-1).

The work of D.~A.~G. was supported by the Foundation for the Development of Theoretical Physics and Mathematics "BASIS" (project No. 23-1-2-52-1).

We also thank Ivan S. Terekhov for meaningful and inspiring discussions on the topic.

%%%%%%%%%%%%%%%%%%%%%%%%%%%%%%%%%%%%%%%%%%%
\section{Conflict of Interest}
%%%%%%%%%%%%%%%%%%%%%%%%%%%%%%%%%%%%%%%%%%%

The authors declare that they have no conflict of interest.
%

%
%%%%%%%%%%%%%%%%%%%%%%%%%%%%%%%%%%%%%%%%%%%%%%%%%%%%%%%%%%%%%%%%%%%%%%%%

\clearpage


\begin{table}[ht]
\caption{Leading-order contribution and interelectronic-interaction corrections to the quadratic Zeeman effect for the binding energy of the valence $2p_{1/2}$ electron in the ground $(1s)^2(2s)^2 2p\,^2P_{1/2}$ state of boron-like ions with the nuclear charges $Z=10$, $18$, and $24$ in terms of the coefficient $g^{(2)}$ defined in (\ref{eq:ebb}). The results of calculations performed both in the Coulomb potential (Coul) and with the inclusion of the core-Hartree (CH) and Kohn-Sham (KS) screening potentials are presented. ${\text{FS}}$ denotes the estimates of the quadratic Zeeman effect given by equations \eqref{eq:nrl:gbb} and \eqref{eq:est-gbb}, {LO} denotes the leading-order contribution \eqref{eq:dirac}, {1ph} denotes the one-photon-exchange correction. The last row shows the sum of the {LO} and {1ph} contributions.}
\label{tab:int}
\setlength{\tabcolsep}{20pt}
\begin{tabular}{lS[table-format=-7.5,group-separator=,table-align-text-post=false]S[table-format=-7.5,group-separator=,table-align-text-post=false]S[table-format=-7.5,group-separator=,table-align-text-post=false]}
\hline
Term&{Coul}&{CH}&{KS}\\
\hline
\multicolumn{4}{c}{ $Z = $ 10}\\
{FS} \eqref{eq:nrl:gbb}&-250770.& {---} & {---}\\
{FS} \eqref{eq:est-gbb}&-249768.&-620567.&-570994.\\
{LO}&-247899.&-617108.&-567585.\\
{1ph}&-208127.&-41904.&-92097.\\
{Sum}&-456025.&-659012.&-659682.\\
\hline
\multicolumn{4}{c}{ $Z = $ 18}\\
{FS} \eqref{eq:nrl:gbb}&-23888.3& {---}& {---} \\
{FS} \eqref{eq:est-gbb}&-23578.3&-37561.3&-36040.7\\
{LO}&-23007.3(00)&-36782.8(00)&-35268.1\\
{1ph}&-10818.7(00)&-1470.4(00)&-2982.8\\
{Sum}&-33826.0(00)&-38253.2(00)&-38250.9\\
\hline
\multicolumn{4}{c}{ $Z = $ 24}\\
{FS} \eqref{eq:nrl:gbb}&-7558.42& {---}& {---} \\
{FS} \eqref{eq:est-gbb}&-7383.60&-10399.53&-10091.16\\
{LO}&-7066.15(00)&-10000.91(00)&-9694.79\\
{1ph}&-2515.02(00)&-298.66(00)&-604.24\\
{Sum}&-9581.18(00)&-10299.57(00)&-10299.03\\
\hline
\end{tabular}
\end{table}

\begin{table}[ht]
\setlength{\extrarowheight}{-1pt} 
\caption{
QED corrections to the quadratic Zeeman effect for the binding energy of the valence $2p_{1/2}$ electron in the ground $(1s)^2(2s)^2 2p,^2P_{1/2}$ state of boron-like ions with the nuclear charges $Z=10$, $18$, and $24$ in terms of the coefficient $g^{(2)}$ defined in (\ref{eq:ebb}). The results of calculations performed both in the Coulomb potential (Coul) and with the inclusion of the core-Hartree (CH) and Kohn-Sham (KS) screening potentials are presented. The total QED correction is also presented in terms of the function $\faz$ defined in (\ref{eq:faz}).}
\setlength{\tabcolsep}{20pt}
\label{tab:QED}
\begin{tabular}{l@{}S[table-format=-7.5,group-separator=,table-align-text-post=false]S[table-format=-7.5,group-separator=,table-align-text-post=false]S[table-format=-7.5,group-separator=,table-align-text-post=false]}
\hline
Term&{Coul}&{CH}&{KS}\\
\hline
\multicolumn{4}{c}{ $Z = $ 10}\\
SE&-590.8(1.0)&-1459.5(1.2)&-1343.5(1.2)\\
VP&0.6&1.4&1.3\\
Sum&-590.2(1.0)&-1458.1(1.2)&-1342.2(1.2)\\
$\faz$&1.017&1.012&1.012\\
\hline
\multicolumn{4}{c}{ $Z = $ 18}\\
SE&-57.52(26)&-90.96(26)&-87.36(26)\\
VP&0.19&0.29&0.28\\
Sum&-57.33(26)&-90.67(26)&-87.08(26)\\
$\faz$&1.046&1.039&1.040\\
\hline
\multicolumn{4}{c}{ $Z = $ 24}\\
SE&-18.50(18)&-25.88(20)&-25.13(21)\\
VP&0.11&0.15&0.15\\
Sum&-18.39(18)&-25.73(20)&-24.98(21)\\
$\faz$&1.072&1.065&1.066\\
\hline
\end{tabular}
\end{table}

\begin{table}[ht]
\setlength{\extrarowheight}{-3pt}
\caption{Individual contributions and total theoretical predictions for the quadratic Zeeman effect for the binding energy of the $2p_{1/2}$ electron in the  $(1s)^2(2s)^2 2p\,^2P_{1/2}$ state of boron-like ions in the range of nuclear charges $Z = 10$--24 in terms of the coefficient $g^{(2)}$ defined in (\ref{eq:ebb}). The results of calculations performed both for the Coulomb potential (Coul) and with the inclusion of the core-Hartree (CH) and Kohn-Sham (KS) screening potentials are presented. LO is the leading-order contribution \eqref{eq:gbb-dirac}; 1ph is the one-photon-exchange correction; QED is the QED correction. The total results are denoted as "Total values".}
\setlength{\tabcolsep}{20pt}
\sisetup{
    separate-uncertainty = false,
    bracket-numbers = true,
    table-format = -6.2,
    table-number-alignment = center-decimal-marker,
    input-open-uncertainty = (,
    input-close-uncertainty = ),
    parse-numbers = true,
    retain-unity-mantissa = false,
}
\label{tab:gbb:final}
\begin{tabular}{lSSS}
\hline
Term&{Coul}&{CH}&{KS}\\
\hline
\multicolumn{4}{c}{ $Z = $ 10}\\
%\hline
{LO}&-247898.6(00)&-617107.8(00)&-567584.9(00)\\
{1ph}&-208126.5(00)&-41904.2(00)&-92097.5(00)\\
{QED}&-590.2(0.7)&-1458.1(1.4)&-1342.3(1.5)\\
{Total value}&-456615.3&-660470.1&-661024.7\\
\hline
\multicolumn{4}{c}{ $Z = $ 12}\\
%\hline
{LO}&-118942.8(00)&-248234.1(00)&-232207.9\\
{1ph}&-83349.4(00)&-14527.3(00)&-30629.4\\
{QED}&-286.0(2)&-591.7(5)&-554.1(5)\\
{Total value}&-202578.2&-263353.1&-263391.4\\
\hline
\multicolumn{4}{c}{ $Z = $ 14}\\
%\hline
{LO}&-63816.12(00)&-118349.86(00)&-111940.25\\
{1ph}&-38403.11(00)&-6024.99(00)&-12440.73\\
{QED}&-155.11(25)&-284.93(11)&-269.84(12)\\
{Total value}&-102374.35&-124659.78&-124650.82\\
\hline
\multicolumn{4}{c}{ $Z = $ 16}\\
%\hline
{LO}&-37147.38(00)&-63296.44(00)&-60338.01\\
{1ph}&-19602.78(00)&-2838.85(00)&-5794.97\\
{QED}&-91.34(30)&-154.11(16)&-147.13(15)\\
{Total value}&-56841.51&-66289.40&-66280.11\\
\hline
\end{tabular}
\end{table}
\begin{table}[ht]
\setlength{\extrarowheight}{-3pt}
\setlength{\tabcolsep}{20pt}
\begin{tabular}{lS[table-format=-6.5,group-separator=,table-align-text-post=false]S[table-format=-6.5,group-separator=,table-align-text-post=false]S[table-format=-6.5,group-separator=,table-align-text-post=false]}
\hline
Term&{Coul}&{CH}&{KS}\\
\hline
\multicolumn{4}{c}{ $Z = $ 18}\\
%\hline
{LO}&-23007.28(00)&-36782.80(00)&-35268.07\\
{1ph}&-10818.70(00)&-1470.44(00)&-2982.81\\
{QED}&-57.33(26)&-90.67(26)&-87.08(26)\\
{Total value}&-33883.31&-38343.92&-38337.95\\
\hline
\multicolumn{4}{c}{ $Z = $ 20}\\
%\hline
{LO}&-14960.93(00)&-22760.38(00)&-21921.35\\
{1ph}&-6349.09(00)&-819.11(00)&-1656.58\\
{QED}&-37.80(23)&-56.86(24)&-54.86(24)\\
{Total value}&-21347.82&-23636.35&-23632.79\\
\hline
\multicolumn{4}{c}{ $Z = $ 24}\\
%\hline
{LO}&-7066.15(00)&-10000.91(00)&-9694.79\\
{1ph}&-2515.02(00)&-298.66(00)&-604.24\\
{QED}&-18.39(18)&-25.73(20)&-24.98(21)\\
{Total value}&-9599.57&-10325.30&-10324.01\\
\hline
\end{tabular}
\end{table}

\end{document}